\documentclass[10pt,journal,cspaper, compsoc]{IEEEtran}

\usepackage{graphicx}
\usepackage{epstopdf}
\usepackage{subfig}
\usepackage{amsmath}
\usepackage{amssymb}
\usepackage{amsfonts,dsfont}
\usepackage{hyperref}
\usepackage{breakurl}
\usepackage{caption}
\usepackage{verbatim}
\usepackage{algorithm}
\usepackage{algorithmic}
\usepackage{color}

\newcommand{\CN}{\mathcal{N}}

\newcommand{\CS}{\mathcal{S}}

\newcommand{\CD}{\mathcal{D}}

\newcommand{\CC}{\mathcal{C}}
\newcommand{\CL}{\mathcal{L}}
\newcommand{\CNU}{\mathcal{NU}}

\newcommand{\red}[1]{\color{black} #1}

\newtheorem{theorem}{{\bf \hspace{-0.18in} Theorem}}

\newtheorem{definition}{{\bf\hspace{-0.18in} Definition}}
\def\done{\hspace*{\fill} \rule{1.8mm}{2.5mm} }

\begin{document}

\title{Friends or Foes:  Distributed and Randomized Algorithms to
    Determine Dishonest Recommenders in Online Social Networks}

\author{Yongkun Li,~\IEEEmembership{Member,~IEEE} \hskip 20pt
        John C.S. Lui,~\IEEEmembership{Fellow,~IEEE}

\IEEEcompsocitemizethanks{\IEEEcompsocthanksitem Yongkun Li is with the
school of computer science and technology, University of Science and
Technology of China.\protect\\ E-mail: yongkunlee@gmail.com
\IEEEcompsocthanksitem John C.S. Lui is with the Department of Computer
Science and Engineering,
The Chinese University of Hong Kong.\protect\\
E-mail: cslui@cse.cuhk.edu.hk
}
\thanks{}}


\IEEEcompsoctitleabstractindextext{
\begin{abstract}
Viral marketing is becoming important due to the popularity of online
social networks (OSNs).  Companies may provide incentives (e.g., via free
samples of a product) to a small group of users in an OSN, and these users
 provide recommendations to their friends, which eventually increases the
 overall sales of a given product. Nevertheless, this also opens a door
for ``{\em malicious behaviors}'': dishonest users may intentionally give
misleading recommendations to their friends so as to distort the normal
sales distribution. In this paper, we propose a detection framework to
identify dishonest users in OSNs. In particular, we present a set of fully
distributed and randomized algorithms, and also quantify the performance
of the algorithms by deriving probability of false positive,  probability
of false negative, and the distribution of number of detection rounds.
Extensive simulations are also carried out to illustrate the impact of
misleading recommendations and the effectiveness of our detection
algorithms. The methodology we present here will enhance the security
level of viral marketing in OSNs.
\end{abstract}


}

\maketitle

\IEEEdisplaynotcompsoctitleabstractindextext

\IEEEpeerreviewmaketitle

\section{ Introduction}\label{sec:introduction}

{ \red \IEEEPARstart{I}{n} the past few years, we have witnessed an exponential growth of user population in online
social networks (OSNs).  Popular OSNs such as Facebook, Twitter and Taobao~\cite{taobao} have attracted millions of
active users. Moreover, due to the rapid development of intelligent cell phones and their integration of online social
networking services \cite{zhang14,zhangtc13},  many  users have integrated these services  into their daily
activities, and they often share various forms of information with each other. For example, users share their opinions
on purchased products with their friends, and they may also receive or even seek recommendations from their friends
before doing any purchase. Therefore, when one buys a product, she may be able to influence her friends to do further
purchases. This type of influence between users in OSNs is called the {\em word-of-mouth effect}, and it is also
referred as {\em social influence}.

Due to the large population and the strong social influence in OSNs,
companies are also adapting a new way to reach
their potential customers. In particular, instead of using the conventional
broadcast-oriented advertisement (e.g., through TV or newspaper), companies
are now using target-oriented advertisement which takes advantage of the social
influence so to attract users in OSNs to do their purchases.
This new form of advertisement can be described as follows:
firms first attract a small fraction of initial users in OSNs
by providing free  or discounted samples, then rely on the
word-of-mouth effect to finally attract a large amount of buyers.
As the word-of-mouth effect spreads quickly in social networks,
this form of advertisement is called the {\em viral marketing}, which is
a proven and effective way to increase the sales and revenue for
companies~\cite{DR01, wordofmouth01, viralmarket06, DR02}.

We like to emphasize that viral marketing in OSNs do exist in the real world.
One particular prime example
is the Taobao~\cite{taobao}, which is one of the major operations
under the Alibaba group,  and it is the
biggest e-commerce website in China.  As of June 2013,
Taobao has over 500 million registered users and 60 million
of regular visitors per day. It also hosts
more than 800 million types of products and represents 100 million
dollars turnover per year \cite{taobaosize, taobaosize2}.
Users can buy various types of products from Taobao, and they can also run
their own shops in selling products. Moreover, Taobao also developed an
application addon called {\em Friends Center} on top of the website.
With this application addon, users in Taobao can follow other users just like
the following relationship in Twitter, hence, an OSN is formed
on top of this e-commerce system.  In this OSN,  users can share
many types of information with their friends, including the products they  purchased,
the shops they visited,  as well as the usage experiences or opinions on
products or shops. In particular,
users can also forward their friends' posts or even give comments. In addition to using
the Friends Center in Taobao, one can also associate  her Taobao account with  her account
in Sina Weibo~\cite{weibo, taobaoassociateweibo}, which is  the biggest OSN in China.
According to Weibo's prospectus, the monthly active users of Weibo reached
143.8 million  in March 2014,
and the daily active users also reached 66.6 million \cite{weibosize}.
By associating Taobao account with Sina Weibo, users can easily share
their purchasing experience and ratings on products with their friends in Weibo.
Based on Taobao and Weibo, many companies can
easily perform target-oriented advertisement to promote their products. In fact,
this type of advertisement can be easily launched in any OSN.
}

However, the possibility of doing target-oriented advertisement in OSNs also
opens a door for malicious activities. Precisely,  dishonest users in an OSN
may intentionally give misleading recommendations to their neighbors, e.g.,
by giving high (low) rating on a low-quality (high-quality) product. To take
advantage of the word-of-mouth effect, firms may also hire some users in an
OSN to promote their products. In fact, this type of advertisement becomes
very common in Taobao and Weibo. Worse yet, companies may even consider paying
users to badmouth their competitors' products. Due to the misleading
recommendations given by dishonest users, even if a product is of low
quality, people may still be misled  to purchase it. Furthermore, products
 of high  quality may lose out since some potential buyers are
diverted to other low-quality products.

As we will show in Section~\ref{sec: simulation impact} via simulation,
misleading recommendations made by dishonest users indeed have a significant
impact on the market share. In particular, a simple strategy of promoting
one's own product while bad-mouthing competitors' products  can greatly
enhance the sales of one's product. Furthermore, even if a product is of low
quality, hiring a small percentage of users to promote it by providing
misleading recommendations can severely shift the market share on various
products. Therefore, it is of big significance to identify  dishonest users
and remove them from the networks so as to maintain the viability of viral
marketing in OSNs. With respect to the normal users in OSNs, it is also of
big interest to identify the dishonest users among their neighbors so as to
obtain more accurate recommendations and  make  wiser decisions on
purchasing products. Motivated by this, this paper addresses the problem of
detecting dishonest recommenders in OSNs, in particular, {\em how can a
normal user discover and identify foes from a set of friends during a
sequence of purchases}?

However, it is not an easy task to accurately identify  dishonest users in
OSNs. First, an OSN usually contains millions of users, and the friendships
among these users are also very complicated, which can be indicated by the
 high clustering coefficient of OSNs. Second, users in an OSN
 interact with their friends very frequently, which makes it
difficult to identify dishonest users by tracing and analyzing the behaviors
of all users in a centralized way. Last but not the least, in the scenario
of OSNs, honest users may also have malicious behaviors unintentionally,
e.g.,  they may simply forward the received misleading recommendations given
by their dishonest neighbors without awareness. Conversely, dishonest users
may also act as honest ones sometimes so as to confuse their neighbors and
try to evade the detection. Therefore, the distinction between dishonest
users and honest ones in terms of their behaviors becomes obscure, which
finally makes the detection more challenging.

To address the problem of identifying dishonest recommenders in OSNs,
this work makes the following contributions:
\begin{itemize}
 \item We propose a fully {\em distributed and randomized algorithm} to
     detect dishonest recommenders in OSNs. In particular, users in an OSN
      can independently execute the algorithm to distinguish their
     dishonest neighbors from honest ones. We further exploit the
     distributed nature of the algorithm by integrating the detection
     results of neighbors so as to speed up the detection, and also extend
     the detection algorithm to handle network dynamics, in particular,
     the ``user churn'' in OSNs.
 \item We provide theoretical analysis on quantifying the performance of
     the detection algorithm, e.g., probability of false positive,
     probability of false negative, and the distribution of time rounds
     needed to detect dishonest users.
 \item We  carry out extensive simulations to validate the accuracy of the
     performance analysis, and further validate the effectiveness of our
     detection algorithm using a real dataset.
\end{itemize}

The outline of this paper is as follows. In Section~\ref{sec: related}, we
review related work and  illustrate the difference  of detecting dishonest
users in OSNs from that in general recommender systems. In Section~\ref{sec:
preliminary}, we   formulate the type of recommendations and the behavior of
users in OSNs. In Section~\ref{sec: detection}, we present the detection
algorithm in detail, and also provide theoretical analysis on the
performance of the algorithm. In Section~\ref{sec:cooperative}, we develop a
cooperative algorithm to speed up the detection, and in
Section~\ref{sec:churn}, we design a scheme to deal with  the network dynamics of
OSNs. We demonstrate the severe impact of misleading recommendations and validate the
effectiveness of the detection algorithms via simulations in
Section~\ref{sec: simulation}, and finally conclude the paper in
Section~\ref{sec: conclusion}.

\section{Related Work}\label{sec: related}
 A lot of  studies focus on the information spreading effect in OSNs,
e.g., \cite{influencemodelKKT03, meanfieldTON}, and  results show that OSNs
are very beneficial for information spreading due to their specific  natures
 such as  high clustering coefficient. To take advantage of the
easy-spreading nature and the large population  of OSNs, viral marketing
which is based on the word-of-mouth effect is becoming popular and has been
widely studied, e.g., \cite{DR01, wordofmouth01, viralmarket06, DR02}. In
particular,  because of the strong social influence in OSNs, a small
fraction of initial buyers  can even attract a large amount of users to
finally purchase the product \cite{meanfieldTON, meanfield09}. A major portion of
viral marketing research thinks viral marketing as an information diffusion process,
and then study the influence maximization problem, e.g., \cite{influencemodelKKT03,Chen2010}. However,
viral marketing in OSNs also opens a door for malicious behaviors as
dishonest recommenders can easily inject misleading recommendations into the
system so to misguide normal users' purchases.

In the aspect of  maintaining system security,  some work like
\cite{spear2007} considers to exploit the framework of trust structure
\cite{TrustCarbone, KrukowJornal, TrustWeeks}.
The rough idea is to compute a trust value for every pair of nodes in
distributed systems. This framework is
 suitable for building delegation systems and reputation systems,
while it still faces a lot of challenges to address the problem of
identifying dishonest users in OSNs studied in this work. First,  OSNs
usually contain millions of users and billions of links, the
 cost on computing the trust values for every pair of users must
be extremely high. Second, even if  the trust values for every pair of users
have been computed, it still requires a mapping from the trust value to a
tag indicating whether a user is dishonest or not, which
 is also a challenging task, especially when users have no priori information
about the number of  dishonest users.

With respect to  malicious behavior detection, it was widely studied in wireless networks (e.g.,
\cite{wirelessdetect09, Theodorakopoulos07, LiTMC13WMCattack}), P2P networks (e.g., \cite{p2pattack05,
performance2010}),  general recommender systems \cite{RecSysSurvey} (e.g., \cite{ShillAttackWIDM05, shilling03,
shillingrecommendersystem}), and online rating systems (e.g., \cite{ArjunKDD13,Fei13,Rahman14}). Unlike previous
works, in this paper we address the problem of malicious behavior detection in a different application scenario,
online social networks. In particular, we focus on the identification of dishonest recommenders in OSNs, which is a
very different problem and also  brings different challenges even comparing to the problems of shill attack detection
in recommender systems and review spam detection in online rating systems, both of which are considered to be more
similar to the problem we considered in this paper.  For example, every user in an OSN may give recommendations to her
friends, while this is totally different from the case of recommender system where recommendations are made only by
the system and are given to users in a centralized way. Second, users' ratings, i.e., the recommendations, may
propagate through the network for OSNs, while there is no recommendation propagation in recommender systems. And this
forwarding behavior makes normal users in OSNs also have the chance of doing malicious activities, e.g., forwarding
neighbors' misleading recommendations without awareness. Last but not the least,  an OSN usually has an extremely
large number of nodes and links and also evolves dynamically, so a distributed detection algorithm becomes necessary
with the consideration of the computation cost. However, this increases  the difficulty of the detection because the
detector only has the local information including her own purchasing experience and the recommendations received from
her neighbors, but not the global information of the whole network as in recommender systems. In terms of the
detection methodology, related work studying  review spam detection usually uses machine leaning techniques, while our
detection framework is based on suspicious set shrinkage with  distributed iterative algorithms.

\section{ Problem Formulation}
\label{sec: preliminary}

In this section, we  first present the model of OSNs and  give the formal
 definitions on different types of recommendations, then we formalize the behaviors of
 users in OSNs
 on how to provide recommendations. In particular, considering that the
objective of dishonest users is to promote their target products and
decrease the chance of being detected, we formalize the behaviors of
dishonest users into a probabilistic strategy.

\subsection{Modeling on Online Social Networks}
We model an OSN as an undirected graph $G=(V,E)$, where $V$ is the set of
nodes in the graph and $E$ is the set of undirected edges. Each node $i \in
V$ represents one user in an OSN, and each link $(i,j) \in E$ indicates the
friendship between  user $i$ and user $j$, i.e.,  user $i$ is a neighbor or friend of
user $j$ and vice versa. That is, user $i$ and user $j$ can interact with
each other via the link $(i,j)$, e.g., give recommendations to each other.
Usually, OSNs are scale-free \cite{meanfield09,meanfieldTON} and the degrees
of nodes follow power law distribution \cite{barabasi99}. Precisely,
$p(k)\propto k^{-\gamma},$ where $p(k)$ is the probability of a randomly
chosen node in $G$ having degree $k$ and $\gamma$ is a constant with a
typical value  $2<\gamma<3$. We denote $\mathcal{N}_i = \{j | (i,j) \in E\}$
as the neighboring set of  user $i$ and  assume that $|\mathcal{N}_i|=N$.

\subsection{Products and Recommendations}
We first formalize products, and then  give the  definitions of different
types of recommendations. We consider a set of ``{\em substitutable}''
products \mbox{$P_1$, $P_2$, $\cdots$, $P_M$} that are produced by firms
$F_1$, $F_2$, $\cdots$, $F_M$, respectively, and these firms compete in the
same market. Two products are substitutable if they are compatible, e.g.,
polo shirts from brand X and brand Y are substitutable goods from the
customers' points of view. We characterize each product $P_j$ with two
properties: (1) its sale price and (2) users' valuations. We assume that
each product $P_j$ has a unique price which is denoted as $p_j$. With
respect to users' valuations, since different users  may have different
ratings on a product because of their subjectivity, we denote $v_{ij}$ as
the valuation of user $i$ on product $P_j$.

We categorize a product into two types according to its  sale price and
users' valuations. In particular, if user $i$ thinks that a product $P_j$ is
sold at the price that truly reveals its quality, then she considers this
product as a {\em trustworthy product}. That is, product $P_j$ is classified
as a trustworthy product by user $i$ only when $p_j = v_{ij}$. Here the
equal sign means that the product is sold at a {\em fair} price  from the
point of view of user $i$. Conversely,  if user $i$ thinks that the price of
product $P_j$ does not reveal its quality, or formally, $p_j \neq v_{ij}$,
then she classifies it as  an {\em untrustworthy product}. Similarly, here
the inequality sign just means that user $i$ thinks that $P_j$ is priced
unfair, maybe much larger than its  value. For example, maybe this product
is of low quality or even bogus, but it is produced by speculative and
dishonest companies  who always seek to maximize their profit by cheating
customers. Formally, we use $T_i(P_j)$ to denote the type of product $P_j$
classified by user $i$, and we have
\begin{equation*}
    T_i(P_j)\!\!=\!\!\left\{\!\!\!
                \begin{array}{ll}
                  1, & \!\!\text{if user $i$ considers $P_j$ to be trustworthy,} \\
                  0, & \!\!\text{if user $i$ considers $P_j$ to be untrustworthy.}
                \end{array}
                \right.
\end{equation*}

Since products are categorized into two types,  we assume that there are two
types of recommendations: positive recommendations and negative
recommendations, which are denoted by $R^P(P_j)$ and $R^N(P_j)$,
respectively.
\begin{definition}
A {\em positive recommendation} on product $P_j$ ($R^P(P_j)$) always claims
   that $P_j$ is a trustworthy product regardless of its
type, while  a {\em negative recommendation} on $P_j$ ($R^N(P_j)$) always
claims that $P_j$ is an untrustworthy product regardless of its type.
Formally, we have
  \begin{eqnarray*}
    R^P(P_j)  &\triangleq&  \text{``$P_j$ is a trustworthy product''},\\
    R^N(P_j)  &\triangleq&  \text{``$P_j$ is an untrustworthy product''}.
  \end{eqnarray*}
\label{def:p_n_rec}
\end{definition}
\vskip -10pt

Note that a recommendation, either $R_P(P_j)$ or $R_N(P_j)$, does {\em not}
reveal the type of  product $P_j$ classified by users, so one may make
positive (or negative) recommendations even if she takes the product as
an untrustworthy (or a trustworthy) product. To have the notion of
correctness, we further classify recommendations into correct
recommendations and wrong recommendations by integrating users' valuations.
\begin{definition}
A recommendation on product $P_j$ is correct for user $i$, which is denoted
as $R_i^C(P_j)$, only when it reveals the  type of $P_j$ classified by user
$i$, i.e., $T_i(P_j)$, while a {\em wrong recommendation} on product $P_j$
for user $i$ ($R_i^W(P_j)$)
 reveals the opposite type of product $P_j$ classified by user $i$. Formally, we have
  \begin{eqnarray*}
    R_i^C(P_j)  \!\!\!\!\!\!&\triangleq&\!\!\!\!\!\!  \left\{ \begin{array}{ll}
                       \!\!\! R^P(P_j),  \!\!&\! \mbox{if $T_i(P_j) \!=\! 1$,}\\
                       \!\!\! R^N(P_j), \!\! &\! \mbox{if $T_i(P_j) \!=\! 0$.}
                      \end{array} \right.\\
    R_i^W(P_j)  \!\!\!\!\!\!&\triangleq& \!\!\!\!\!\! \left\{\begin{array}{ll}
                       \!\!\! R^P(P_j),  \!\!& \!\mbox{if $T_i(P_j) \!=\! 0$,}\\
                       \!\!\! R^N(P_j),  \!\!&\! \mbox{if $T_i(P_j) \!=\! 1$.}
                       \end{array}  \right.
  \end{eqnarray*}
\end{definition}

\subsection{Behaviors of Users in OSNs}
In this subsection, we formalize the behaviors of users in an OSN. We assume
that for any user, if she buys a product, then she can valuate the product
based on her usage experience, and then categorizes it into either a
trustworthy product or an untrustworthy product from her point of view.

{\bf Behaviors of honest users.} We define honest users as the ones who will
not intentionally give wrong recommendations. That is, if an honest user
buys a product, since she can valuate the product and determine its type,
she always gives correct recommendations on the  product to her neighbors.
Precisely, she gives  positive recommendations if the product is considered
to be trustworthy and negative recommendations otherwise.

{\red
On the other hand, if an honest user did not buy a product,  she may also
give recommendations to her neighbors by simply forwarding the received
recommendations from  other ones.
This type of forwarding behavior is quite common in OSNs.
For example, in Taobao and Weibo,
many users forward their friends' posts, including their  purchasing experiences
and ratings on products.
In a study of online social networks~\cite{forwarding_support}, it was
found that 25\% of users had forwarded
an advertisement to other users in the network.
Moreover, due to the anonymity of users' identities (e.g.,
users usually  use pseudo names to register their Weibo account),
it is extremely difficult to trace the information spreading process in OSNs, and
so users may simply forward any
form of information without confirming its truthfulness.
In particular, a user may forward a positive (negative) recommendation given by
her neighbors without validating the quality of the product.
Because of this forwarding behavior, it is possible that honest users  may give
wrong recommendations to their neighbors. Thus,  a user
who gives wrong recommendations is not {\em strictly} dishonest, but only
{\em potentially} dishonest. In other words, if the detector considers a
product to be trustworthy and receives negative recommendations from a neighbor,
she still can not be certain that this neighbor is dishonest,
mainly because it is possible that this neighbor does not intend to cheat, but
is just misled by her neighbors.
}

{\bf Behaviors of dishonest users.} We define dishonest users as the ones
who may give wrong recommendations intentionally, e.g., give positive
recommendations on an untrustworthy product. Note that
 dishonest users may also behave differently as they may aim for promoting
different products, e.g., users who are hired by firm $F_i$ aim for
promoting product $P_i$, while users who are hired by firm $F_j$ aim for
promoting $P_j$. Without loss of generality, we assume that there are $m$
types of dishonest users who are hired by firms $F_1$, $F_2$, $\cdots$, $F_m$,
and they  promote products $P_1$, $P_2$, $\cdots$, $P_m$, respectively.
{\red
Furthermore, we assume that the products promoted by dishonest users (i.e.,
products $P_1$, $P_2$, $\cdots$, $P_m$) are  untrustworthy  for all users.
The intuition  is that these products are of low quality (or even bogus)
so that they can be easily identified by people. The main reason to make
this assumption is that  in this case dishonest users have  incentives
to promote these products for a larger profit,
meanwhile, honest users also have incentives to detect such dishonest users
so as to avoid purchasing untrustworthy products. To further illustrate this,
note that when users in an OSN are attracted to buy a product promoted by
dishonest users, if the product is a trustworthy one, then there is
no difference for these buyers to purchase other trustworthy products instead
of the one promoted by dishonest users,
and so honest users have no incentive to identify the dishonest users who promote
trustworthy products.
In other words, promoting trustworthy products can be regarded as
normal behaviors, so we only focus on the case where the promoted products
 are untrustworthy in this paper. However, we would like to
point out that when we model
the behaviors of dishonest users in the following, we  allow dishonest users
to behave as honest ones and give correct recommendations.
}

Recall that the goal of dishonest users is to attract  as many users as
possible to purchase the product they  promote,  one simple and intuitive
strategy  to achieve this goal is to give positive recommendations on the
product they promote and negative recommendations on all other products.
On the other hand, besides attracting as many users as possible to buy their
promoted product, dishonest users also hope to avoid being detected so that
they can perform malicious activities for a long time. Therefore, dishonest
users may also adopt a more intelligent strategy so to confuse the detector
and decrease the chance of being detected. For instance, instead of always
bad-mouthing other products by giving negative recommendations, they may
{\em probabilistically} give correct recommendations and behave like honest
users sometimes. The benefit of this probabilistic strategy is to make the
detection more difficult so that dishonest users may hide in a longer time.
In this paper, we allow dishonest users to adopt  this intelligent strategy
and use $S_j^l$ to denote the one adopted by a type-$l$ ($1\leq l \leq m$)
dishonest user $j$. Moreover, we allow dishonest users to be more powerful
by  assuming that they
 know honest users' valuation on each product so that they can mislead as many
users as possible.
The intelligent strategy $S_j^l$ can be formally expressed as follows.
\begin{equation}
S_j^l\!\triangleq\! R^P(P_l) \!\wedge\! \Big{[}\!
        \wedge_{n=1, n\neq l}^M
       \left[\delta R_j^C(P_n) \vee (1\!-\!\delta) R^N(P_n)
      \right]\Big{]},
\label{equation: intelligent attack}
\end{equation}
where $\delta$ denotes the probability of  giving correct recommendations.
Recall that the goal of type-$l$ dishonest users is to attract as many users
as possible to purchase  product $P_l$, while giving positive
recommendations on other products (say $P_n, n\neq l$) goes against their
objective, so we assume that dishonest users  only give correct
recommendations on $P_n$ with a small probability, i.e., $\delta$ is small.
In particular,  $\delta=0$ implies that dishonest users always bad-mouth
other products.

{\red
Note that there is a possibility that  dishonest users do not adopt the
probabilistic strategy as in Equation~(\ref{equation: intelligent attack}), while
choose to promote trustworthy products over a long time just to create a good
reputation, and then behave maliciously by giving misleading recommendations on
a product.  However, as long as the dishonest users start performing malicious
activities, our detection framework still provides us with
the opportunity of detecting them  as we can keep executing
the detection algorithm continuously. On the other hand,
even if our framework may fail to detect the dishonest users if they only
perform malicious activities in a very limited number of rounds,
the effect of misleading recommendations
given by dishonest users is also very limited, and so the corresponding miss
detection error should be very small.

Another possibility we would like to point out is that
multiple dishonest users may  {\em collude} and a single dishonest user
may also create multiple Sybil accounts to consistently promote a low-quality product.
This type of collaborated malicious attack is still detectable under our framework,
this is because our detection framework is fully distributed and when the detector
determines whether a neighbor is dishonest or not, she only
relies on her own valuation on a product and the recommendation given by this neighbor.
Therefore, the possibility of a dishonest user being detected only depends on the
amount of malicious activities she performs, and it is irrelevant to other users'
behaviors. However, for the cooperative detection algorithm that is developed
for speeding up the detection, dishonest users may evade the detection if they collude
as the detector may determine the type of a neighbor by exploiting other
neighbors' detection information, while the possibility
of evading the detection depends on the parameters controlled by the detector.
Hence, there is a tradeoff between detection accuracy and  detection efficiency, and
we will further illustrate this in Section~\ref{sec:cooperative}.
}

\subsection{Problem}
In this paper, we develop distributed algorithms that can be run at any user
 in an OSN  to identify  her dishonest  neighbors.  Specifically,
 we first develop a randomized baseline algorithm which only exploits the
 information of the detector,
 see Section~\ref{sec: detection} for details. We also quantify the performance of the algorithm
 via theoretical analysis. Then we propose a cooperative
 algorithm which further takes advantage of the detection results  of the detector's
 neighbors so as to speed up the detection, see Section~\ref{sec:cooperative} for
 details. After that, we  further extend the algorithm to deal with  network dynamics of OSNs,
 i.e., user churn,
 in Section~\ref{sec:churn}.

\section{ Baseline Detection Algorithm}\label{sec: detection}
In this section, we first illustrate the rough idea of the detection
framework, and then  present the detection algorithm in detail. We also
quantify various performance measures of the algorithm.

\subsection{General Detection Framework}\label{subsec:general_framework}

{\red
Our detection algorithm is fully distributed, and so users can independently execute it
to identify dishonest users among their neighbors. Without loss of generality, we only
focus on one particular user, say user $i$, and call
her {\em the detector}. That is, we present the algorithm from
the perspective of user $i$
and discuss how to detect her dishonest neighbors. For  ease of presentation, we
simply call a product as a trustworthy (or untrustworthy) product if
the detector considers it to be trustworthy (or untrustworthy).

Note that even if users' subjectivity creates different
preferences on different products, we assume that the detector and her
neighbors have a consistent valuation  on most products. This
assumption is reasonable, especially for the cases where the quality of
products can be easily identified, and its rationality can be further justified
as follows.  First, users in an OSN prefer to have friends with others who share
similar interests and tastes.
Hence users in an OSN are {\em similar} to their neighbors~\cite{Crandall08}
and so they have a consistent valuation with their neighbors on many products.
Secondly, ``{\em wisdom of the crowd}'' is  considered to be
the basis of online rating systems like Amazon and Epinions, and it
is also widely used by people in their daily lives, so it is reasonable to
assume that most products have intrinsic quality so that
the detector and her neighbors will have a consistent rating.

Note that the above assumption allows users who are not friends with each other
to have very different valuations
on the same product, and it also allows the detector and her neighbors to
have different valuations on some products.
In fact, if an honest neighbor has a different rating on a product,
then from the detector's point of view, it is just equivalent to the case
where this honest neighbor
is misled by dishonest users and so
gives a wrong recommendation.
Another issue we would like to point out
is that even if the above assumption does not hold, e.g.,
if the detector and her neighbors have different ratings on all products,
our detection framework still provides a
{\em significant step toward identifying dishonest behavior in OSNs' advertisement}.
This is because if a neighbor has different valuations on all
products from the detector, then our detection framework will  take this neighbor
as ``misleading'' no matter she intends to cheat or not. This is acceptable
as users always prefer neighbors to have the similar taste (or preference) with
them so that they can purchase a product they really like if they take their
neighbors' recommendations.
}

We model the purchase experience of detector $i$ as a discrete time process.
Particularly, we take the duration between  two continuous purchases made by
detector $i$ as one round, and time proceeds in rounds $t=1,2,\cdots$. That
is, round $t$ is defined as the duration from the time right before the
$t^{th}$ purchase instance to the time right before the $(t+1)^{th}$
purchase instance. Based on this definition, detector $i$ purchases only one
product at each round, while she may  receive various recommendations on the
product from her neighbors, e.g., some neighbors may give her positive
recommendations and others may give her negative recommendations.

{\red
The general idea of our detection framework can be illustrated as in Figure~\ref{fig:general_alg}.
Initially, detector $i$ is conservative and considers {\em all} her
neighbors as {\em potentially} dishonest users. We use $\CS_i(t)$ to denote
the set of potentially dishonest neighbors of detector $i$ until round $t$,
which is termed as {\em the suspicious set}, and  we have
$\mathcal{S}_i(0)=\mathcal{N}_i$. As time proceeds, detector $i$
differentiates her neighbors based on their behaviors in each round, and
shrinks the suspicious set  by removing her trusted neighbors which are
classified as honest users. After sufficient number of rounds, one can
expect that all honest neighbors are removed from the suspicious set and
only dishonest neighbors left. Therefore, after $t$ rounds, detector $i$
takes a neighbor as dishonest if and only if this neighbor belongs to the
suspicious set $\mathcal{S}_i(t)$.

\begin{figure}[!ht]
\begin{center}
\includegraphics[width=0.99\linewidth]{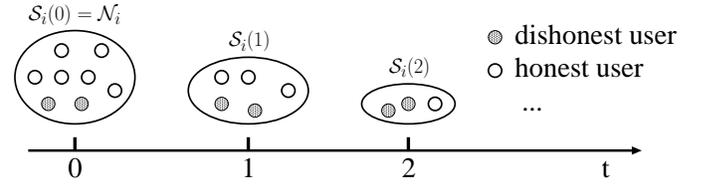}
\caption{General detection process  via  suspicious set shrinkage.}
\label{fig:general_alg}
\end{center}
\end{figure}
}

\subsection{Operations in One Round}\label{subsec:algoneround}
In this subsection, we describe the detailed operations of shrinking the
suspicious set in only one round, say round $t$. Note that detector $i$ buys
a product at round $t$, which we denote as $P_{j_t}$ ($j_t\in
\{1,2,\cdots,M\}$), so she can valuate the product and determine its type
$T_i(P_{j_t})$ from her point of view. Moreover, she can further categorize
the received recommendations (that are either positive or negative) into
correct recommendations and wrong recommendations based on her valuation on
the product, and so she can differentiate her neighbors according to their
recommendations. Specifically, we define $\CN_i^C(t)$ as the set of
neighbors whose recommendations given at round $t$ are classified as correct
 by detector $i$. Accordingly, we denote $\CN_i^W(t)$ and
$\CN_i^N(t)$ as the set of neighbors who give detector $i$ wrong
recommendations and no recommendation at round $t$, respectively. We have
$\CN_i=\CN_i^C(t)\cup\CN_i^W(t)\cup\CN_i^N(t)$.

Recall that a product is either trustworthy  or untrustworthy, so detector
$i$ faces two cases at round $t$: (1) the purchased product $P_{j_t}$ is an
untrustworthy product, i.e., $T_i(P_{j_t})=0$,  and (2) the purchased
product $P_{j_t}$ is a trustworthy product, i.e., $T_i(P_{j_t})=1$. In the
following, we illustrate on how to shrink the suspicious set in the above
two cases.

In the first case, a neighbor who gives correct recommendations can not be
certainly identified as  honest, mainly because a dishonest neighbor may
also give correct recommendations. For example, a type-$l$ ($l\neq j_t$)
dishonest user may give negative recommendations on product $P_{j_t}$ based
on the intelligent strategy, and this recommendation will be classified as
correct  as the detector  valuates product $P_{j_t}$ as an untrustworthy
product. Therefore, detector $i$ is not able to differentiate honest
neighbors from dishonest ones  if $T_i(P_{j_t})=0$. We adopt a conservative
policy by keeping the suspicious set  unchanged, i.e., $\CS_i(t) =
\CS_i(t-1)$.

In the second case, since detector $i$ valuates  product $P_{j_t}$ as  a
trustworthy product,  $P_{j_t}$ cannot be
 a product promoted by dishonest users, and we have $P_{j_t} \in \{P_{m+1}, ...,
P_M\}$. In this case, even if a dishonest user may  give correct
recommendations on product $P_{j_t}$ based on the intelligent strategy, the
corresponding probability  $\delta$ is considered to be small, and so a
dishonest user should belong to the set $N_i^W(t)$ with high probability.
Note that it is also possible that dishonest users do not make any
recommendation at round $t$, so dishonest users can be in either
$\CN_i^W(t)$ or $\CN_i^N(t)$. We use $\CD(t)$ to denote the union of the two
sets, i.e., $\CD(t)=\CN_i^W(t)\cup \CN_i^N(t)$, which denotes the set to
which dishonest users belong with high probability at round $t$. To balance
the tradeoff between  detection accuracy and detection rate, we employ a
{\em randomized policy} that only shrinks the suspicious set with
probability $p$. Precisely, we let $\CS_i(t) = \CS_i(t-1) \cap \CD(t)$ only
with probability $p$. Here $p$ is a tunable parameter chosen by detector
$i$, and it reflects the degree of conservatism of the detector. The
detailed algorithm which is referred as {\em the randomized detection
algorithm} at round $t$ is stated in Algorithm~\ref{alg:baseline}.

\begin{algorithm}
\caption{Randomized Detection Algorithm at Round $t$ for Detector $i$}
\label{alg:baseline}
\begin{algorithmic}[1]
    \STATE Estimate the type of the purchased product $P_{j_t}$;
    \STATE Differentiate neighbors by determining $\CN_i^C(t)$, $\CN_i^W(t)$, and
    $\CN_i^N(t)$;
    \STATE Let $\CD(t) \leftarrow \CN_i^W(t)\cap \CN_i^N(t)$;
    \IF{$T_i(P_{j_t})=1$}
        \STATE with probability $p$: $\CS_i(t) \leftarrow \CS_i(t-1) \cap \CD(t);$
        \STATE with probability $1-p$: $\CS_i(t) \leftarrow \CS_i(t-1);$
    \ELSE
        \STATE $\CS_i(t) \leftarrow \CS_i(t-1);$
    \ENDIF
\end{algorithmic}
\end{algorithm}

{\red
To  further illustrate the detection process in Algorithm~\ref{alg:baseline},
we consider Figure~\ref{fig:baseline_alg} as an example to show the  operations
at round $t$. User $i$ have seven neighbors that are labeled from $a$ to $g$.
Assume that two of them are dishonest (i.e., user $a$ and $b$).
Suppose that neighbors $a$, $b$, $c$ and $e$ are still in the
suspicious set before round $t$,
i.e., $\mathcal{S}_i(t-1)=\{a,b,c,e\}$. We use dashed cycles
to denote  suspicious users in Figure~\ref{fig:baseline_alg}. If user $i$
buys a trustworthy product at round $t$, and only neighbors $e$ and  $f$ give her correct
recommendations, then user $i$ can be certain that neighbor $e$ is honest
with a high probability and it can be removed from the suspicious set.
Therefore, according to Algorithm~\ref{alg:baseline}, the suspicious set shrinks
with probability $p$, and if this probabilistic even happens,
then we have $\mathcal{S}_i(t)=\{a,b,c\}$ as shown
on the right hand side of Figure~\ref{fig:baseline_alg}.

\begin{figure}[!ht]
\begin{center}
\includegraphics[width=0.99\linewidth]{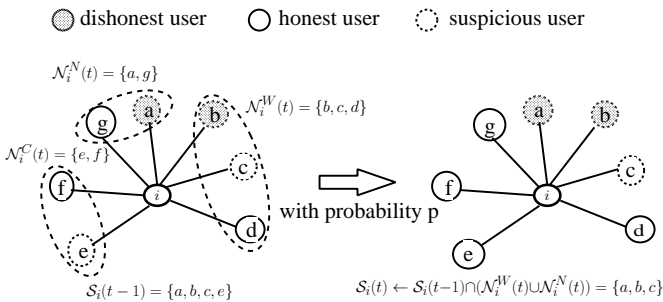}
\caption{An example illustrating Algorithm~\ref{alg:baseline}.}
\label{fig:baseline_alg}
\end{center}
\end{figure}
}

 Note that Algorithm~\ref{alg:baseline} is fully distributed in
the sense that it can be executed by any user to identify  her dishonest
neighbors. The benefit of the distributed nature is twofold. First, the size
of an OSN is usually very large, e.g., it may contain millions of nodes and
billions of links, so a distributed algorithm becomes necessary so as to
make the computation feasible. Second, an OSN itself is fully distributed,
in particular, a user in an OSN only receives information, e.g.,
recommendations on products, from her direct neighbors, and so she only
needs to care about the honesty of her neighbors so as to make the received
recommendations more accurate. Therefore, a fully distributed detection
algorithm is indeed necessary for the application we consider.

In terms of the implementation of Algorithm~\ref{alg:baseline},  it can be
deployed as a third-party application just like others that are deployed in
OSNs. In particular, when this application has been deployed in an OSN, each
user has a choice to install it or not. If a user chooses to install it,
then she needs to submit some necessary information to the social network
provider continuously, e.g., her ratings on products and the recommendations
that she would like to make, and the provider will aggregate and store the
information for each user. Finally, the computation can be done by either
the server of the social network provider or the client computer of each
user.

\subsection{Performance Evaluation}\label{subsec:alg1performance}

To characterize the performance of the detection algorithm, we define three
performance measures: (1) {\em probability of false negative} which
 is  denoted as
$P_{fn}(t)$,
 (2) {\em probability of false positive} which  is denoted  as
$P_{fp}(t)$, and (3) {\em the  number of rounds needed to shrink the
suspicious set until it only contains dishonest users}, which is denoted by
a random variable $R$. Specifically,  $P_{fn}(t)$ characterizes the
probability that a dishonest user is wrongly regarded as an honest one after
$t$ rounds, and $P_{fp}(t)$ characterizes the error that an honest user is
wrongly regarded as a dishonest one after $t$ rounds. Recall that detector
$i$ takes a neighbor $j \in \mathcal{N}_i$ as dishonest if and only if this
neighbor belongs to the suspicious set (i.e., $j \in \mathcal{S}_i(t)$), so
we define $P_{fn}(t)$ as the probability that a dishonest neighbor of
detector $i$ is not in $\mathcal{S}_i(t)$ after $t$ rounds. Formally, we
have
\begin{equation}
\begin{small}
    P_{fn}(t)  = \frac{\text{\# of dishonest neighbors of $i$ that are not in } \mathcal{S}_i
(t)}
    {\text{total \# of dishonest neighbors of detector } i}.
    \label{eq:pfndef}
\end{small}
\end{equation}
On the other hand,  since all neighbors of detector $i$ are initially
included in the suspicious set (i.e., $\mathcal{S}_i(0)=\mathcal{N}_i$), an
honest user is wrongly regarded as a dishonest one only if she still remains
in the suspicious set after $t$ rounds. Thus, we  define $P_{fp}(t)$ as the
probability of an honest user not being removed from the suspicious set
after $t$ rounds. Formally, we have
\begin{equation}
    P_{fp}(t) = \frac{\text{\# of honest neighbors of $i$ that are in }
      \mathcal{S}_i(t) }{\text{total \# of honest neighbors of detector } i}.
    \label{eq:pfpdef}
\end{equation}

To derive the above three performance measures for
Algorithm~\ref{alg:baseline}, note that the suspicious set shrinks at round
$t$ only when detector $i$ valuates her purchased product as  a trustworthy
product and  this round is further used for detection with probability $p$.
We call such a round a {\em detectable} round and use a 0-1 random variable
$d(t)$ as an indicator, where $d(t)=1$ means that round $t$ is detectable
and 0 otherwise. In addition to the indicator $d(t)$, detector $i$  also
obtains the set $\CD(t)$ to which dishonest users may belong  at round $t$.
Therefore, we use a tuple $(d(t), \CD(t))$ to denote the information that
detector $i$ obtains at round $t$, and  the set of all tuples until round
$t$ constitute the {\em detection history}, which we denote as
$\mathcal{H}(t)$. Formally, we have
\begin{equation*}
  \mathcal{H}(t) = \{(d(1), \CD(1)),(d(2), \CD(2)),...,(d(t), \CD(t))\}.
\end{equation*}

Based on the detection history $\mathcal{H}(t)$, the performance measures of
$P_{fn}(t)$, $P_{fp}(t)$ and the distribution of $R$ for
Algorithm~\ref{alg:baseline} can be
 derived as
in Theorem \ref{theo:randomized}.

\begin{theorem}
{\em After running Algorithm \ref{alg:baseline} for $t$ rounds, probability
of false negative and probability of false positive are derived in
Equation~(\ref{eq:pfn2}) and Equation~(\ref{eq:pfp}), respectively.
\begin{eqnarray}
P_{fn}(t)&=&1-\left(1-\delta\right)^{\sum_{\tau=1}^{t}{d(\tau)}},    \label{eq:pfn2}\\
P_{fp}(t)&\approx& \prod_{\tau=1, d(\tau)=1}^{t}{\frac{|\CD(\tau-1)\cap
\CD(\tau)|}{|\CD(\tau-1)|}}, \label{eq:pfp}
\end{eqnarray}
where $\CD(0)=\CN_{i}$ and $\CD(\tau)$ is set as $\CD(\tau-1)$ if $d(\tau)=0$.

\noindent The number of rounds needed for detection until the suspicious set only
contains dishonest users follows the distribution of
\begin{eqnarray}
P(R\!=\!r)\!\!\!\!\!\!&=&\!\!\!\!\!\!\sum_{d=1}^{r}\!{\binom{r-1}{d-1}(p_d)^d(1\!-\!p_d)^{r-d}}
\times\nonumber\\
\!\!\!\!\!\!&&\!\!\!\!\!\!\big{[}[1\!-\!(1\!-\!p_{hc})^d]^{N-k} \!\!-\!\! [1\!-\!(1\!-\!p_{hc})^
{d-1}]^{N-k}\big{]},
    \label{eq:R}
\end{eqnarray}
where $p_{hc}$ is the average probability of an honest user giving correct
recommendations at each round and  $p_d$ is the probability of a round being
detectable, which can be  estimated by Equation~(\ref{eq:p_hc}) and
Equation~(\ref{eq:p_d}) in the Appendix, respectively.}
\label{theo:randomized}
\end{theorem}
\noindent {\bf Proof:} Please refer to the Appendix. \done

{\red
Since  probability of false positive $P_{fp}(t)$ is critical to
design the complete detection algorithm (see Section~\ref{subsec:complete}),
we use an example to further illustrate its derivation.
Note that a user in an OSN usually has a large number of friends, so we let
detector $i$ have 100 neighbors  labeled from 1 to 100.
Among these 100 neighbors, we assume that the last two are dishonest, whose labels
are  99 and  100. Before starting the detection algorithm,
we initialize $\CD(0)$ as $\CN_{i}$ and let $P_{fp}(0)=1.$

In the first detection round, suppose that user $i$ buys a trustworthy product and
further takes this round as detectable. Besides, suppose that
only neighbor 1 and neighbor 2 give her correct recommendations,
i.e., $\CD(1)=\{3,4,\cdots,100\}$,
then we have $\CS_{i}(1)=\{3,4,\cdots,100\}$.
Based on Equation~(\ref{eq:pfp}), the probability of false positive
can be derived as $$P_{fp}(1)=P_{fp}(0)*\frac{|\CD(0)\cap \CD(1)|}{|\CD(0)|}=0.98.$$
Note that according to the definition in Equation~(\ref{eq:pfpdef}),
the accurate value of  probability of false positive is $\frac{96}{98}$, which
is a little bit smaller than the result derived by Theorem~\ref{theo:randomized}.
In fact, Theorem~\ref{theo:randomized} provides a good approximation when
the number of neighbors is large and the number of dishonest users among them is small,
which is the common case for OSNs as users often  tend to have
a lot of friends and  a company can only control a small number of users to promote
its product.

Now let us consider the second detection round.
Suppose that the event with probability $p$
does not happen. That is,  this round is not detectable. So
we set $\CD(2)=\CD(1)$, and the suspicious set
remains the same, i.e., $\CS_{i}(2)=\CS_{i}(1)=\{3,4,\cdots,100\}$.
The probability of false positive is still $$P_{fp}(2)=0.98.$$

We further examine one more round.  Suppose that the third round
is detectable and  neighbor 1 to neighbor 4 give user $i$ correct recommendations,
i.e., $\CD(3)=\{5,\cdots,100\}$. Based on Algorithm~\ref{alg:baseline},
we have $\CS_{i}(3)=\CS_{i}(2)\cap \CD(3)=\{5,\cdots,100\}$. The probability of
false positive can be derived as
$$P_{fp}(3)=P_{fp}(2)*\frac{|\CD(2)\cap \CD(3)|}{|\CD(2)|}=0.96.$$
Note that according to the definition in Equation~(\ref{eq:pfpdef}),
the accurate value after round $t$ is
$\frac{94}{98}=0.959$.
}

Based on Theorem~\ref{theo:randomized}, we see   that $P_{fp}(t)\rightarrow 0$,
and this implies
that all honest users will be removed from the suspicious set  eventually.
However, $P_{fn}(t)$ does not converge to zero, which implies that
dishonest users may evade the detection. Fortunately, as long as $P_{fn}(t)$ is
not too large when $P_{fp}(t)$ converges to zero, one can still effectively
identify {\em all}  dishonest users (as we will show in Section \ref{sec:
simulation}) by executing the detection process multiple times.
On the other hand, the expectation of $R$ quantifies the {\em efficiency} of the detection
algorithm, in particular,  it indicates how long a detector needs to
identify her dishonest neighbors on average. Note that the detection algorithm itself
does not rely on the derivation of this performance measure,
 and it is just used for studying the detection efficiency of the algorithm.

\subsection{{\bf Complete Detection Algorithm}} \label{subsec:complete}
In Section~\ref{subsec:algoneround}, we present a partial detection
algorithm which describes the operations in  a particular  round $t$. In
this subsection, we present the corresponding complete algorithm which
describes how to shrink the suspicious set until dishonest users can be
identified. To achieve this, we have to determine the termination condition
when repeating the partial algorithm round by round. Observe that after
executing the detection algorithm for $t$ rounds, only users in the
suspicious set $\mathcal{S}_i(t)$ are taken as dishonest ones. Intuitively,
to avoid a big detection error, the detection process can only be terminated
when users in $\mathcal{S}_i(t)$ are really dishonest with {\em high
probability}. Based on the definition of probability of false positive
$P_{fp}(t)$,  it is sufficient to terminate the algorithm when $P_{fp}(t)$
is lower than a predefined small threshold $P_{fp}^*$. In other words, as
long as probability of false positive is small enough, we can guarantee that
all users in the suspicious set are really dishonest with high probability.
Based on the above illustration, the complete detection algorithm  can be
stated as follows.
\begin{algorithm}
\caption{Complete Detection Algorithm} \label{alg:baselinecomplete}
\begin{algorithmic}[1]
    \STATE $t\leftarrow 0$;
    \STATE $\CS_i(0)\leftarrow\CN_i$;
    \REPEAT
        \STATE $t\leftarrow t+1$;
        \STATE Derive the suspicious set $\CS_i(t)$ at round $t$ by executing Algorithm \ref
{alg:baseline};
        \STATE Update  probability of  false positive $P_{fp}(t)$;
    \UNTIL{$P_{fp}(t) \leq P_{fp}^*$}
    \STATE Take users in  $\mathcal{S}_i(t)$ as dishonest and blacklist them;
\end{algorithmic}
\end{algorithm}

\section{ Cooperative Algorithm to Speed up the Detection}\label{sec:cooperative}
In the last section, we propose a distributed and randomized algorithm that
only exploits the detector's local information. By running this algorithm,
honest users can detect their dishonest neighbors simultaneously and
independently. That is, each user in an OSN maintains her own suspicious set
containing her potentially dishonest neighbors. Since users in an OSN
interact with each other frequently,  they can also share their detection
results, e.g., their suspicious sets. By doing this, a detector can further
exploit her neighbors' detection history  to speed up her own detection, and
we term this scenario as {\em cooperative detection}.

We still focus on a particular detector, say user $i$, and use $\CS_i(t)$ to
denote her suspicious set. At round $t$, user $i$ may shrink her suspicious
set based on her purchasing experience and her received recommendations, and
she may also request the detection results of her neighbors. In particular,
we assume that detector $i$ can obtain two sets from each neighbor $j$ at
round $t$: the neighboring set and the suspicious set of neighbor $j$, which
we denote as $\mathcal{N}_j$ and $\mathcal{S}_j(t)$, respectively.

To exploit neighbors' detection results, at round $t$, detector $i$ first
shrinks her own suspicious set according to Algorithm~\ref{alg:baseline}, and we call this
step as {\em the independent detection step}. After that, detector $i$
 further shrinks her  suspicious set by exploiting the information
received from her neighbors (i.e., $\{(\mathcal{N}_j, \mathcal{S}_j(t)),\:
j \in \mathcal{N}_i\}$), and we term this step as {\em the cooperative detection
step}. Since detector $i$ may have different degrees of
trust on her neighbors, we use $w_{ij}(t)$ ($0\leq w_{ij}(t) \leq 1$) to
denote the weight of trust of user $i$ on neighbor $j$ at round $t$. That
is,  user $i$ only exploits the detection results  of  neighbor $j$  with
probability $w_{ij}(t)$ at round $t$. Intuitively, $w_{ij}(t)=1$ implies
that user $i$ fully trusts neighbor $j$, while $w_{ij}(t)=0$ means that
user $i$ does not trust $j$ at all. The cooperative detection algorithm for
user $i$ at round $t$ is stated in Algorithm~\ref{alg:cooperative}.

\begin{algorithm}
\caption{Cooperative Detection Algorithm at Round $t$ for Detector $i$}
\label{alg:cooperative}
\begin{algorithmic}[1]
   \STATE Derive the suspicious set $\mathcal{S}_i(t)$  based on local information (i.e.,  using
Algorithm \ref{alg:baseline});
    \STATE Exchange detection results with neighbors;
   \FOR{each neighbor $j\in \mathcal{N}_i$}
        \STATE with probability $w_{ij}(t)$: $\mathcal{S}_i(t) \leftarrow \mathcal{S}_i(t)
\backslash (\mathcal{N}_j \backslash \mathcal{S}_j(t))$;
        \STATE with probability $1-w_{ij}(t)$: $\mathcal{S}_i(t) \leftarrow \mathcal{S}_i(t)$;
   \ENDFOR
\end{algorithmic}
\end{algorithm}

{\red
We take Figure~\ref{fig:cooperative_alg} as an example to
further illustrate the operations at the cooperative detection step
(i.e., Line~3-6 in Algorithm~\ref{alg:cooperative}).
Since user $i$ first shrinks her suspicious set by using
Algorithm~\ref{alg:baseline}, we still use the setting in
Figure~\ref{fig:baseline_alg} where $\mathcal{S}_{i}(t)$ shrinks
to $\{a,b,c\}$ after the first step. Now to further exploit neighbors'
detection results to shrink $\mathcal{S}_{i}(t)$, suppose that only user $c$
is a neighbor of user $d$, and it has already been removed from user $d$'s
suspicious set. That is, $c \in \CN_{d}$ and  $c \notin \CS_{d}$. If
user $i$ fully trusts neighbor $d$ (i.e., $w_{id}(t)=1$),
then user $i$ can be  certain that neighbor $c$ is honest as $c$ is
identified as honest by neighbor $d$.
Thus, user $i$ can further shrink her suspicious set, and we have
$\mathcal{S}_{i}(t)=\{a,b\}$ as shown
on the right hand side of Figure~\ref{fig:cooperative_alg}.

\begin{figure}[!ht]
\begin{center}
\includegraphics[width=0.99\linewidth]{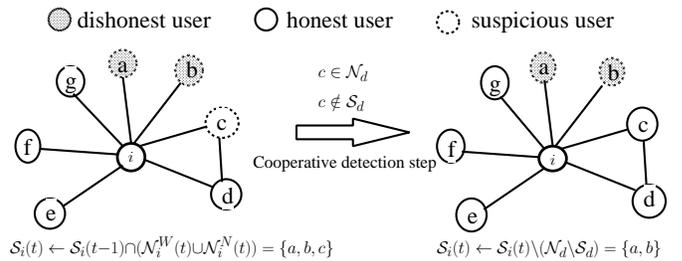}
\caption{An example illustrating Algorithm~\ref{alg:cooperative}.}
\label{fig:cooperative_alg}
\end{center}
\end{figure}

}

To implement Algorithm~\ref{alg:cooperative}, we need to set the weights of trust on
different neighbors, i.e., $w_{ij}(t)$. One simple  strategy is
 only trusting the neighbors that are not in the suspicious set as users in the suspicious set are
potentially dishonest.
 Mathematically,
we can express this strategy as follows.
\begin{equation}
    w_{ij}(t) = \left\{
    \begin{array}{cc}
      0, & \textrm{if} \:j \in \mathcal{S}_i(t), \\
      1, & \textrm{otherwise.}
    \end{array}
    \right.
\end{equation}
{\red

Note that $w_{ij}(t)$ is a tunable parameter for \mbox{detector $i$}, and
it affects the shrinking rate of the suspicious set of detector $i$.
On the other hand, since detector $i$ may further
shrink her suspicious
set by exploiting her neighbors' detection results,
dishonest users may evade the detection if they collude, while the possibility also
depends on the parameter $w_{ij}(t)$.  In fact,    there is
a tradeoff between detection accuracy and efficiency when choosing this
parameter.  Specifically,  larger $w_{ij}(t)$'s
imply that detector $i$ is more aggressive to further exploit her neighbors'
detection results, and so the detection rate should be larger, while the risk of
dishonest users evading the detection also becomes  larger.
}

Again, Algorithm \ref{alg:cooperative} is only a partial algorithm that
describes the operation at round $t$. To develop the complete version of the
cooperative detection algorithm, we can still use the idea in
Section~\ref{subsec:complete} to set the termination condition. That is, we
keep running Algorithm~\ref{alg:cooperative} until  probability of false
positive is less than a predefined threshold $P_{fp}^*$. To achieve this, we
have to derive the probability of false positive $P_{fp}(t)$ for
Algorithm~\ref{alg:cooperative}, and the result is stated in
Theorem~\ref{theo:cooperative}.

\begin{theorem}
{\em After running Algorithm \ref{alg:cooperative} for $t$ rounds,
 probability of false positive can be derived as follows.
\begin{equation*}
P_{fp}(t)\approx \frac{P_{fp}(t-1)\frac{|\CD(t-1)\cap \CD(t)|}{|\CD(t-1)|} N -|\CC(t)|} {N},
\end{equation*}
\label{theo:cooperative} }
\end{theorem}
where  $P_{fp}(0) = 1$ and $\CC(t)$  denotes the set of neighbors
that are removed from the suspicious set in the cooperative detection step
at round $t$.

 \noindent {\bf Proof:} Please refer to the Appendix.
\done

\section{Algorithm Dealing with User Churn} \label{sec:churn}
In previous sections, we proposed a randomized detection algorithm and also
discussed about how to speed up the detection. These algorithms are designed
based on the assumption that the underlying network is static, i.e., the
friendships between users are fixed and do not change during the detection.
 However, an online social network usually evolves dynamically,
in particular, new users may join in the network and existing users may
change their friendships or even leave the network by deleting their
profiles \cite{Golbeck07,Leskovec2008,Zhao2012}. Taking the the network
dynamics into consideration, for detector $i$, new users may become her
friends and existing friends may also disconnect with her at some time. We
call these behaviors as {\em user churn}.  Note that even if users may leave
the network and rejoin it after some time, while they may not be able to
recover the past friendships as establishing links or friendships usually
requires the confirmation of other users in OSNs. In this section, we extend
our detection algorithm to address the problem of user churn in OSNs.

We still focus on a particular detector, say user $i$. At each round,  we
first employ previous algorithms, e.g., Algorithm \ref{alg:baseline} or
Algorithm \ref{alg:cooperative}, to shrink the suspicious set. After that,
we do the following checks: (1) whether there are new users becoming the
neighbors of detector $i$, and (2) whether some existing neighbors of
detector $i$ disconnect with her. In particular, if new neighbors come in,
we add them into the neighboring set $\mathcal{N}_i$ and the suspicious set
$\mathcal{S}_i(t)$. In other words,  we are
conservative to take new users  as potentially dishonest. For ease of
presentation, we use $\CNU(t)$ to denote the set of  {\em new users} that
become the neighbors of detector $i$ at round $t$. On the other hand, if
some existing neighbors disconnect with detector $i$ at round $t$, we simply
remove them from both the neighboring set $\mathcal{N}_i$ and the suspicious
set $\mathcal{S}_i(t)$. We  use $\CL(t)$ to denote the set of neighbors that
{\em leave} detector $i$ at round $t$, and use $\CL_S(t)$ to denote
the set of users that are in the suspicious set $\mathcal{S}_i(t)$ and leave detector $i$
at round $t$,
 i.e., $\CL_S(t) =
\mathcal{S}_i(t) \cap \CL(t)$. Now we present the detailed detection
algorithm at round $t$ in Algorithm~\ref{alg:userchurn}. Note that if
Algorithm \ref{alg:cooperative} is used to shrink the suspicious set in
Algorithm \ref{alg:userchurn}, then cooperative detection is  used to speed
up the detection.
\begin{algorithm}
\caption{Dealing with User Churn at Round $t$} \label{alg:userchurn}
\begin{algorithmic}[1]
    \STATE Derive the suspicious set $\mathcal{S}_i(t)$ (by executing
    Algorithm \ref{alg:baseline} or Algorithm \ref{alg:cooperative});
    \STATE Derive the set $\CNU(t)$ and $\CL(t)$;
    \STATE $\mathcal{S}_i(t) \leftarrow (\mathcal{S}_i(t) \cup \CNU(t)) \backslash \CL(t)$;
    \STATE $\mathcal{N}_i \leftarrow (\mathcal{N}_i \cup \CNU(t)) \backslash \CL(t)$;
\end{algorithmic}
\end{algorithm}

{\red
Let us use an example to illustrate the operations in Algorithm~\ref{alg:userchurn}
and it is shown in Figure~\ref{fig:userchurn_alg}. Since  the suspicious set
first shrinks by using Algorithm~\ref{alg:baseline} or Algorithm~\ref{alg:cooperative},
which has been illustrated before. Here we only show the step dealing with user churn
(i.e., Line~2-4). Suppose that at round $t$, user $i$ disconnects with
neighbor $b$ (i.e., $\CL(t)=\{b\}$), and initiates
a connection with a new user that is labeled as $h$ (i.e., $\mathcal{NU}(t)=\{h\}$),
then user $i$ can safely remove $b$ from the suspicious set as she does not
care user $b$ any more, while she has no priori information about the type
of the new user $h$, so she is conservative and add user $h$ into the suspicious set.
Thus, we have $\mathcal{S}_{i}(t)=\{a,h\}$ as shown in Figure~\ref{fig:userchurn_alg}.

\begin{figure}[!ht]
\begin{center}
\includegraphics[width=0.99\linewidth]{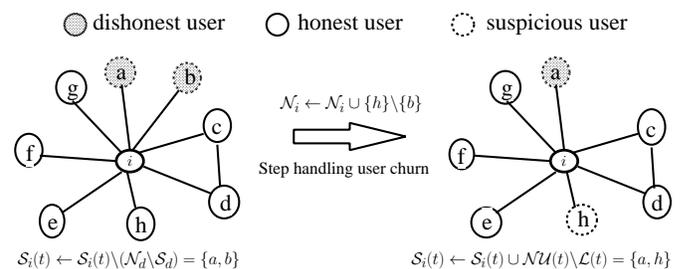}
\caption{An example illustrating Algorithm~\ref{alg:userchurn}.}
\label{fig:userchurn_alg}
\end{center}
\end{figure}

}

The  complete algorithm can also be developed by keeping
running the detection process until probability of false positive is smaller
than a predefined threshold $P_{fp}^*$. Thus,  we have to derive the
probability of false positive $P_{fp}(t)$ for Algorithm \ref{alg:userchurn},
and the result is stated in Theorem~\ref{theo:userchurn}.
\begin{theorem}
{\em After running Algorithm~\ref{alg:userchurn} for $t$ rounds,
 probability of false positive can be derived as follows.
 {\small
\begin{equation*}
P_{fp}(t)\!\!\approx\!\!\frac{P_{fp}(t\!\!-\!\!1)\frac{|\CD(t\!-\!1)\cap \CD(t)|}{|\CD(t\!-\!
1)|}
                        N(t\!-\!1)\!\!-\!\!|\CC(t)| \!\!+\!\! |\CNU(t)|\!\! -\!\! |\CL_S(t)|}
{N(t)},\nonumber
\end{equation*}
}
\label{theo:userchurn} }
\end{theorem}
where  $P_{fp}(0) = 1$ and $N(t)$ denotes the number of neighbors after round $t$.

\noindent {\bf Proof:} Please refer to the Appendix. \done

\section{ Simulation and Model Validation}\label{sec: simulation}
Our model aims to detect dishonest users who intentionally give wrong
recommendations in OSNs. Since each user in an OSN performs her own
activities continuously, e.g., purchasing a product, giving recommendations
to her neighbors, and making decisions on which product to purchase, the
network evolves dynamically. Therefore, we first synthesize a dynamically
evolving social network to emulate users'  behaviors, then we show the
impact of misleading recommendations and  validate the analysis  of our
detection algorithm based on the synthetic network. We also validate the
effectiveness of our detection algorithm using a real dataset drawn from an
online rating network.

\subsection{ Synthesizing A Dynamically Evolving OSN}\label{subsec:systemevolve}
In this subsection, we synthesize a dynamic OSN to simulate the behaviors of
users in the network. To achieve this, we make assumptions on (1) how users
make recommendations to their neighbors, (2) how users make decisions on
purchasing which product,
 and (3) how fast the recommendations  spread.

First,  there are two types of users in the network: honest users and
dishonest users. Dishonest users adopt the intelligent strategy to make
recommendations. For  an honest user, if she buys a product, she  gives
correct recommendations to her friends based on her valuation on the
product. On the other hand, even if an honest user does not buy a product,
she  still gives recommendations based on her received recommendations. We
adopt the majority rule in this case. That is, if more than half of her
neighbors give positive (negative) recommendations to her,  then she gives
positive (negative) recommendations to others. Otherwise, she does not give
any recommendation. In the simulation, we let all honest users have the same
valuation on each product, and so we  randomly choose an honest user as the
detector in each simulation.

Second, to simulate the behaviors of users on deciding to purchase which
product, we assume that an honest user  buys the product  with the maximum
number of effective recommendations that is defined as the number of
positive recommendations subtracting the number of negative recommendations.
The rationale is that one buys a product that receives high ratings  as many
as possible and low ratings  as few as possible.

Last, we assume that the spreading rate of recommendations is much higher
than  the purchasing rate. In other words, when one gives a positive
(negative) recommendation on a particular product to her neighbors, her
neighbors update their states accordingly, i.e., update the number of
received positive (negative) recommendations. If the corresponding numbers
satisfy the majority rule, then they  further make recommendations on this
product, and this process continues until no one in the system  can make a
recommendation according to the majority rule. Moreover, the whole process
finishes before the next purchase instance made by any user in the network.

To model the evolution of the network, we assume that it starts from the
``uniform'' state in which all products have the same market share. During
one detection round,  $10\%|V|$ purchase instances happen, where $|V|$ is
the total number of users in the network, i.e., between two successive
purchases of detector $i$, $10\%|V|$ purchases are made by other users in
the network. Note that the assumptions we make in this subsection are only
for the simulation purpose, and our detection algorithms do not require
these assumptions.

\subsection{ Impact of Misleading Recommendations} \label{sec: simulation impact}
In this subsection, we  show the impact of misleading recommendations using
the synthetic network. We employ the GLP model proposed in \cite{GLP} that
is based on preferential attachment \cite{barabasi99} to generate a
scale-free graph with power law degree distribution and high clustering
coefficient. We generate a graph with around 8,000 nodes and 70,000 edges,
whose  clustering coefficient is around 0.3. We assume that initially no
product has been purchased,  and consider 10,000 purchase instances in the
simulation. For each purchase instance, one user purchases and she buys the
product with the maximum number of effective recommendations. After that,
she gives a recommendation on the product to her friends. The recommendation
will spread throughout the network until no one can make a recommendation
according to the majority rule. We assume that there are five products,
$P_1, \cdots, P_5$, and dishonest users aim to promote product $P_1$ which
is an untrustworthy product, while the rest are trustworthy products. Our
objective is to measure the fraction of purchases of each product out of the
total 10,000 purchases. We run the simulation multiple times and take the
average value.
\begin{figure}[!htb]
  \centering
  \includegraphics[width=0.7\linewidth]{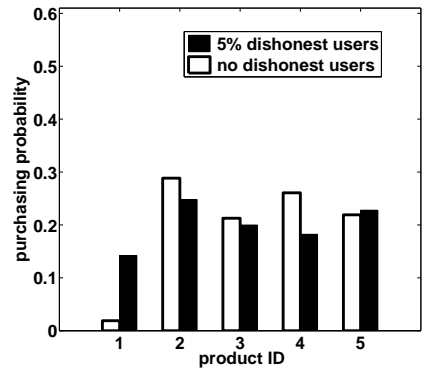}
  \caption{Impact of misleading recommendations on the market share distribution:
  dishonest users aim to promote an untrustworthy product $P_1$.}
  \label{fig:effect}
\end{figure}

 The simulation results are shown
in Figure \ref{fig:effect}. First, we can see that if  no dishonest user
exists in the network  to give misleading recommendations, the untrustworthy
product $P_1$ is purchased with only a small probability. The reason why the
probability is non-zero is that if a user does not receive any
recommendation, she  simply makes a random choice over the five products to
make a purchase. However, if we randomly set 5\% of users as dishonest and
let them adopt the intelligent strategy to promote $P_1$ by setting
$\delta=0$, then even if $P_1$ is an untrustworthy product, it is still
purchased with probability around 0.15. In other words, many users in the
network are misled by these dishonest users to purchase $P_1$. In summary,
the existence of dishonest users who   intentionally give misleading
recommendations can severely distort the  market share distribution.

\subsection{ Analysis Validation via A Synthetical OSN}
In this subsection, we  synthesize a dynamically evolving network based on
the description in Section \ref{subsec:systemevolve}, and then validate our
analysis on the performance of  the detection algorithms. In the simulation,
we
 randomly select 5\% of users as dishonest users, and let them  adopt the
  intelligent strategy. We also randomly
choose an honest user who has dishonest neighbors  and take her as the
detector. We carry out the simulation many times and take the average value
as the simulation results.

\begin{figure}[!ht]
  \centering
  \includegraphics[width=0.7\linewidth]{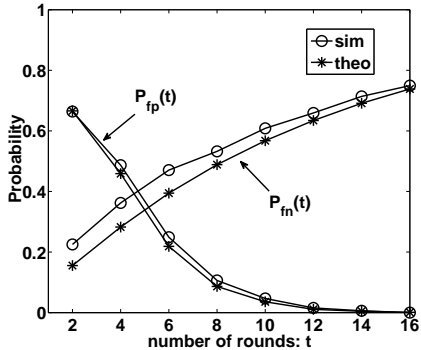}
  \caption{Probability of false negative and probability of false positive of the randomized
detection algorithm
   (Algorithm~\ref{alg:baseline})
           where $\delta=0.1$ and $p=0.8$.}
  \label{fig:A2_pfnp}
\end{figure}
Let us first focus on the performance measures of $P_{fn}(t)$ and
$P_{fp}(t)$ for Algorithm~\ref{alg:baseline}. The theoretic results and
simulation results are shown in Figure \ref{fig:A2_pfnp}. First, we can see
that the theoretic results match well with the simulation results. Second,
one only needs to run the detection algorithm for a small number of rounds
to remove all honest users from the suspicious set, which shows the
effectiveness and efficiency of the detection algorithm. However,
probability of false negative is not zero as dishonest users may act as
honest ones sometimes with the hope of evading the detection. This implies
that only a part of dishonest users are detected  in one execution of the
algorithm. Fortunately, when probability of false positive goes to zero,
probability of false negative is still not close to one. Therefore, to
detect all dishonest users, one can run the
 algorithm multiple times. At each time, a subset of
dishonest users are detected and then removed. Eventually, all dishonest
users can be identified. For example, in Figure \ref{fig:A2_pfnp}, after ten
rounds, probability of false positive is close to zero, and probability of
false negative is just around 0.6, which indicates that at least 40\% of
dishonest users can be detected in one execution of the algorithm.

\begin{figure}[!ht]
  \centering
  \includegraphics[width=0.7\linewidth]{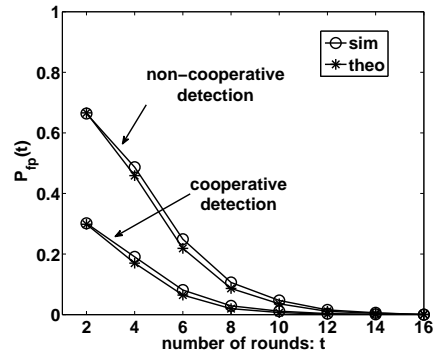}
  \caption{The improvement of probability of false positive for the cooperative algorithm
(Algorithm \ref{alg:cooperative})
  where $\delta=0.1$ and $p=0.8$. }
  \label{fig:A4_pfp}
\end{figure}

Now we focus on the cooperative detection algorithm, i.e., Algorithm
\ref{alg:cooperative}. Figure \ref{fig:A4_pfp} compares the probability of
false positive for the randomized detection algorithm (Algorithm
\ref{alg:baseline}) with its corresponding cooperative version (Algorithm
\ref{alg:cooperative}). Results show that our theoretic analysis provides a
good approximation of probability of false positive, which validates the
effectiveness of  the termination condition used in the complete detection
algorithm. Moreover,
 comparing the two groups of curves, we can see that
probability of false positive of the cooperative algorithm is always smaller
than that of the non-cooperative algorithm, which implies that the
cooperative scheme effectively speeds up the detection.

\begin{figure}[!htb]
  \centering
  \includegraphics[width=0.7\linewidth]{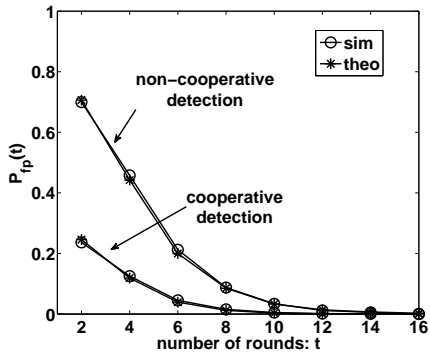}
  \caption{Probability of false positive of the algorithm dealing with user churn (Algorithm
\ref{alg:userchurn})
  where $\delta=0.1$ and $p=0.8$. }
  \label{fig:A6_pfp}
\end{figure}

Now we focus on  the detection algorithm dealing with user churn, i.e.,
Algorithm \ref{alg:userchurn}, and the results are   shown in Figure
\ref{fig:A6_pfp}. In the figure, one group of curves corresponds to the case
where cooperative algorithm is employed, i.e., using Algorithm
\ref{alg:cooperative} to derive the suspicious set in the first step of
Algorithm \ref{alg:userchurn}, the other  group corresponds to the case
where cooperative detection is not used, i.e., using Algorithm
\ref{alg:baseline} to derive the suspicious set in the first step. To
simulate user churn, we add a new neighbor to the detector with probability
0.3 in each round. Simulation results show that probability of false
positive goes to zero eventually, which implies that users in the suspicious
set must be dishonest with high probability after sufficient number of
rounds. At last, we also observe the speedup of the detection for the
cooperative  algorithm.

\begin{figure}[!htb]
  \centering
  \includegraphics[width=0.7\linewidth]{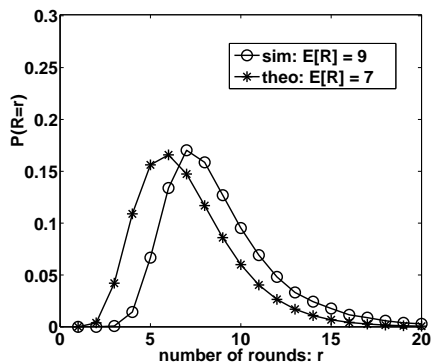}
  \caption{Probability mass function of $R$ when the randomized detection algorithm is used and
$\delta=0.1$ and $p=0.8$.}
  \label{fig:distR}
\end{figure}

Let us  look at the distribution of  number of detection rounds for the
randomized detection algorithm, i.e., Algorithm~\ref{alg:baseline}. Results
are shown in Figure \ref{fig:distR}. The horizontal axis is the number of
rounds needed for the detection, and the vertical axis is the probability
mass function. We can see that even if the probability mass function is not
accurately quantified, the expected number of rounds, $E[R]$, is  still well
approximated. The deviation  of the probability mass function  can be
explained as follows. First, the probability of an honest user giving
correct recommendations is not a constant at each round, e.g., as more users
purchase a product, the probability of giving correct recommendations also
increases since more users can have their own valuations. Therefore, there
must be an approximation error when we use a constant parameter, say
$p_{hc}$, to approximate it. Second, since the performance measure is
quantified in a probabilistic way, it is required to run the simulation many
times so as to match with the theoretic results. However, running the
simulation too many times takes a lot of time because of the large graph
size. To balance the tradeoff, we only run the simulation 1000 times, and
the inadequate number of simulation times   also contributes to the
approximation error. However, since  the detection algorithm does not
require the accurate quantification of the distribution of  $R$, it is still
effective to employ the algorithm to identify dishonest users even if an
approximation error exists.

\subsection{Evaluation on Real Data}

As we stated in Section~\ref{sec:introduction}, the problem we considered in
this paper
is an abstraction
of viral marketing problems in OSNs,
and so there is no publicly available
dataset that is drawn from an OSN specialized for viral marketing.
Therefore, to further validate the effectiveness of our detection algorithm,
we consider a real dataset from a social rating network, where users share
their ratings on movies and also establish friendships with  others. In the
following, we first describe the dataset, then illustrate on how to
implement our detection algorithm, and finally show the  results.

{\bf Dataset}: We use the Flixster dataset which is drawn from a social
network where users share their ratings on movies with their friends
\cite{Jamali2010}. The underlying social network contains around 1M users
and 26.7M social relations. Users can give ratings in the range $[0.5,5]$
with step size 0.5, and there are  8.2M ratings in this dataset. Since we
classify products into two types in this work, i.e., trustworthy products
and untrustworthy products, to drive evaluations using the Flixster dataset,
we map each rating to a binary value (i.e., either 0 or 1) by splitting from
the middle of the range (i.e., 2.5). That is, we take the ratings that are
greater than 2.5 as high ratings, and consider others as low ratings. By
analyzing this dataset, we find that for around 90\% of movies, more than
75\% of users have a consistent valuation. This also confirms the assumptions we make
in our framework.

{\bf Algorithm Implementation}: Since our detection algorithm is fully
distributed and can be executed by any user. To select a detector, we randomly
choose a user who has a large number of friends and also gives a lot of
ratings. In particular, the detector chosen in this evaluation has around
900 friends and gives around 200 ratings. Among the  900 friends, around 700
of them have only one rating or even no rating on all of the movies  the
detector rated, so we ignore them in the evaluation. Since all users in this
dataset are honest, to emulate malicious activities, we randomly set 10\% of the
 detector's neighbors as dishonest users and let them promote one particular
 movie. In particular, we modify the  ratings given by these dishonest users
 based on the intelligent strategy formalized in Equation~(\ref{equation: intelligent attack}).
 We run the randomized detection algorithm (i.e.,
 Algorithm~\ref{alg:baseline}) at the detector, and measure the probability
 of false negative and the probability of false positive based on the
 definitions
 in Equations~(\ref{eq:pfndef})-(\ref{eq:pfpdef}) so as to validate the effectiveness of the
detection algorithm.

\begin{figure}[!ht]
  \centering
  \includegraphics[width=0.7\linewidth]{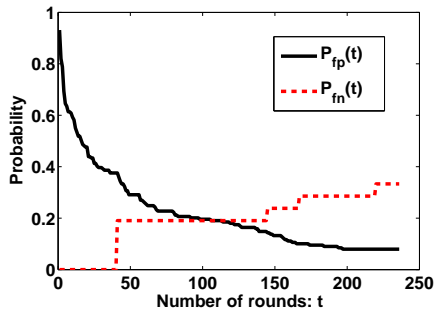}
  \caption{Probability of false positive and probability of false negative of
  the randomized detection algorithm on real dataset. }
  \label{fig:real}
\end{figure}

 {\bf Detection Results}: The results of  probability of false positive $P_{fp}(t)$ and
  probability of false negative $P_{fn}(t)$ are shown in
 Figure~\ref{fig:real}. We can see that probability of false positive
 continues to decrease as the algorithm executes for more and more rounds, and finally
 falls below a small probability. This implies that most honest users can be
 successfully removed from the suspicious set. On the other hand, probability
 of false negative also increases, which indicates the possibility of miss detection.
 However, we can see that when probability of false
 positive drops below 0.1, probability of false negative  only increases to
 0.3. This shows the effectiveness of the detection algorithm. In particular,
 more than 70\% of dishonest users can be accurately identified in
 one execution of the algorithm, and so we can keep executing the algorithm
 for multiple times so to identify all dishonest users. Another important point we like
 to stress is that the number of detection rounds in this evaluation
 is not small, e.g., probability of false positive only decreases to 0.3 after
 50 rounds. The main reason is that in this dataset, most users only give very few
 ratings, and so many neighbors do not give any rating in most of the detection
 rounds, which makes them remain in the suspicious set for a long time.

\section{Conclusion}\label{sec: conclusion}

In this paper, we develop a set of fully distributed and randomized
detection algorithms based on the idea of shrinking suspicious set
 so to identify dishonest users in OSNs.
 We formalize the behaviors of dishonest users wherein
they can probabilistically bad-mouth other products while give positive
recommendations on the product they aim to promote.  Our detection
algorithms allow   users to independently perform the detection so as to
discover their dishonest neighbors. We provide  mathematical analysis on
quantifying the effectiveness and efficiency of the detection algorithms. We
also propose a cooperative scheme  to speed up the detection, as well as an
algorithm to handle network dynamics, i.e., ``user churn'' in OSNs. Via
simulations,  we first show that the  market share distribution may be
severely distorted by  misleading recommendations given by a small fraction
of dishonest users, and then validate the effectiveness and efficiency of
our detection algorithms. The detection framework in this paper can be
viewed as a valuable tool to maintain the viability of viral marketing in
OSNs.

\appendix
\section*{Proof of Theorem~\ref{theo:randomized} in Section~\ref{subsec:alg1performance}}

We first focus on probability of false negative $P_{fn}(t)$. Note that
$\mathcal{S}_i(t)$ only shrinks in detectable rounds,
 so we have
\begin{eqnarray*}
    P_{fn}\!(t)\!\! \!\!\!\!&=&\!\!\!\!\!\! P\{\textrm{\!a dishonest user $j$  is considered to
be honest} \}\\
    \!\!\!\!&=&\!\!\!\!\!\!1 \!-\! P\{\textrm{$j$ is not removed from the suspicious set} \\
    \!\!\!\!&&\!\!\!\!\!\!\!\!\!\!\textrm{\hspace{0.4in} in all detectable rounds}\} \\
   \!\!\!\! &=&\!\!\!\!\!\!1\!-\!\!\!\!\!\!\!\prod_{\tau=1,d(\tau)=1}^{t}\!\!\!\!\!\!\!
               P\{j\in \CD(\tau)\}.
\end{eqnarray*}
To compute the probability that a dishonest user stays in $\CD(\tau)$ in a
detectable round $\tau$, observe that the product  detector $i$ purchases at
this round must be a trustworthy product which is not  promoted by any
dishonest user. We assume that dishonest users give recommendations at every
round so as to attract as many buyers as possible. Based on the intelligent
strategy, a dishonest user gives correct recommendations on this product
with probability $\delta$, and this recommendation is also correct for
detector $i$ as we assume that dishonest users have the same valuation with
majority users,  so we have $P\{j \in \CD(\tau)\} = 1 - \delta$, and
probability of false negative is
\begin{equation*}
    P_{fn}(t) \!=\! 1\!-\!\!\!\!\!\!\!\prod_{\tau=1, d(\tau)=1}^{t}\!\!\!\!\!\!\!(1-\delta)
    =1-\Big{(}1-\delta\Big{)}^{\sum_{\tau=1}^{t}{d(\tau)}}.
\end{equation*}

To derive probability of false positive $P_{fp}(t)$, based on the definition
in Equation~(\ref{eq:pfpdef}), it can be rewritten as
\begin{eqnarray}
  P_{fp}(t)\!\!\!\! &=& \!\!\!\! P\{j \in \mathcal{S}_i(t) | \mathcal{H}(t)\} \nonumber\\
   \!\!\!\! &=&\!\!\!\!  P\{j \in \mathcal{S}_i(t-1) | \mathcal{H}(t)\} \times \nonumber \\
   \!\!\!\! &&\!\!\!\!  P\{j \in \mathcal{S}_i(t) | j \in \mathcal{S}_i(t-1) \& \mathcal{H}(t)\}
\nonumber\\
   \!\!\!\! &=&\!\!\!\! P_{fp}(t-1)\times P\{j \textrm{ is not removed}\nonumber\\
   \!\!\!\!&&\!\!\!\! \textrm{  at round }t | j \in \mathcal{S}_i(t-1) \& \mathcal{H}(t)\},
   \label{eq:pfptemp}
\end{eqnarray}
where $j$ is an honest friend of detector $i$. To compute the probability
that $j$ is not removed at round $t$, we first consider the case where round
$t$ is detectable. Considering that  a user in an OSN
usually has a large number of neighbors and dishonest users  only account for a
small fraction,  so we  approximate the probability of an honest user in the
suspicious set not
being removed at round $t$ as $\frac{|\CD(t-1)\cap \CD(t)|}{|\CD(t-1)|}$.
On the other hand,
if round $t$ is not
detectable, then the corresponding probability is simply one.
For
ease of presentation, we always let $\CD(t)=\CD(t-1)$ if round $t$ is
not  detectable, so the probability can still be expressed as
$\frac{|\CD(t-1)\cap \CD(t)|}{|\CD(t-1)|}$.
By substituting it in Equation~(\ref{eq:pfptemp}),
 we have
\begin{equation*}
P_{fp}(t) \approx \prod_{\tau=1,d(\tau)=1}^{t}{\frac{|\CD(\tau-1)\cap \CD(\tau)|}{|\CD
(\tau-1)|}},
\end{equation*}
where $\CD(0)$ is initialized as $\mathcal{N}_i$ and
$\CD(\tau)$ is set as $\CD(\tau-1)$ if $d(\tau)=0$.

Now we focus on the third performance measure $R$, which denotes the number
of rounds needed to shrink the suspicious set until it only contains
dishonest users. Note that the suspicious set can shrink at round $t$  only
when this round is detectable, i.e., $d(t)=1$. Therefore, we first derive
the distribution of number of detectable rounds, and denote it  by a random
variable $D$. Formally, we have
\begin{eqnarray*}
  P(D\leq d) \!\!\!\!\!&=& \!\!\!\!\!P\{ \textrm{after $d$ detectable rounds, all users}
\nonumber \\
  \!\!\!\!\!&&\!\!\!\!\!\textrm{ in the suspicious set are dishonest}\}\nonumber \\
   \!\!\!\!\!&=&\!\!\!\!\! P\{ \textrm{all honest users are removed from the}\nonumber \\
    \!\!\!\!\!&&\!\!\!\!\!\textrm{ suspicious set after $d$ detectable rounds}\}\nonumber\\
   \!\!\!\!\!&=&\!\!\!\!\!(1-(1-p_{hc})^d)^{N-k},
\end{eqnarray*}
where $k$ is the number of dishonest neighbors  of detector $i$ and $p_{hc}$
denotes the average probability of an honest user being removed from the
suspicious set at each round, i.e., the average probability of an honest
user giving correct recommendations at each round.

Based on the distribution of $D$, we can derive the distribution of $R$.
Specifically, the conditional distribution $P(R=r|D=d)$ is a negative
binomial distribution, so we have
\begin{eqnarray*}
    \!\!\!\!\!\!\!\!P(R\!=\!r)\!\!\!\!\!\!&=&\!\!\!\!\!\!\!\sum\nolimits_{d=1}^{r}{P(D=d)P(R=r|
D=d)}\nonumber \\
    \!\!\!\!\!\!&=&\!\!\!\!\!\!\!\sum\nolimits_{d=1}^{r}\!{\binom{r-1}{d-1}(p_d)^d(1-p_d)^{r-d}P
(D=d)},
\end{eqnarray*}
where $p_d$ denotes the probability of a round being detectable. Based on
the distribution of $R$, the expected number of rounds $E[R]$ can be easily
derived.

For  probabilities $p_{hc}$ and $p_d$, we can estimate them based on the
detection history of detector $i$. Specifically, to measure $p_{hc}$, for
each honest neighbor $j$ of detector $i$,  we first count the number of
rounds where user $j$ gives correct recommendations to detector $i$, then
use the fraction of rounds where user $j$ gives correct recommendations as
an approximation of the corresponding average probability. Finally, we can
approximate $p_{hc}$ by taking an average over all honest neighbors of
detector $i$. Mathematically, we have
\begin{equation}
{\small
p_{hc}\approx\frac{1}{N-k}\sum_{\text{honest } j \in \CN_i}\!\!\!\!\!\!\!\!
\frac{\text{\# of rounds where $j$ gives correct rec.}}{\text{total \# of rounds}},
\label{eq:p_hc}
}
\end{equation}
where $k$ denotes the number of dishonest neighbors of detector $i$. With
respect to $p_{d}$, note that $p_d$ equals to  the probability that detector
$i$ valuates her purchased product as  a trustworthy product in a round and
this round is further used for detection.  To estimate it, we use a 0-1
random variable $\mathbf{1}\{T_i(P_{j_t})=1\}$ to indicate whether product
$P_{j_t}$ that is purchased by detector $i$ at round $t$ is a trustworthy
product or not, and we have
\begin{equation}
p_d=p\cdot\lim_{n\rightarrow \infty}{\frac{\sum_{t=1}^n
\mathbf{1}\{T_i(P_{j_t})=1\}}{n}}.
\label{eq:p_d}
\end{equation}

\section*{Proof of Theorem~\ref{theo:cooperative} in Section~\ref{sec:cooperative}}

Note that the detection at round $t$ is divided  into
two steps,  the independent detection step and
the cooperative detection step. In the independent detection
step, detector $i$ shrinks her suspicious set based on her local
information, i.e., via Algorithm \ref{alg:baseline}. In the cooperative
detection step, detector $i$ further shrinks her suspicious set based on her
neighbors' detection results. We use $\CC(t)$ to denote the set of neighbors
which are removed from the suspicious set in the cooperative detection step
at round $t$. Based on Equation (\ref{eq:pfptemp}), probability of false
positive can be expressed as follows.
\begin{eqnarray*}
  P_{fp}(t) \!\!\!\! &=&\!\!\!\! P_{fp}(t-1)\times P\{j \textrm{ is not removed}\\
   \!\!\!\!&&\!\!\!\! \textrm{  at round }t | j \in \mathcal{S}_i(t-1) \& \mathcal{H}(t)\}\\
\!\!\!\! &=&\!\!\!\!P_{fp}(t-1) P_{IS}(t) P_{CS}(t),
\end{eqnarray*}
where $j$ an honest friend of detector $i$,  $P_{IS}(t)$ and $P_{CS}(t)$ denote
the probabilities that an honest user in the suspicious set is not removed    in   the
independent detection step and the cooperative detection step,
respectively.


To derive $P_{IS}(t)$, since the suspicious set shrinks based on Algorithm
\ref{alg:baseline} in the independent detection step, we can directly use
the result in Theorem \ref{theo:randomized}. We have
\begin{equation*}
P_{IS}(t) \approx \frac{|\CD(t-1)\cap \CD(t)|}{|\CD(t-1)|},
\end{equation*}
where $\CD(t)$ is set as $\CD(t-1)$ if round $t$ is not detectable (i.e., when $d(t) = 0$).

To compute $P_{CS}(t)$ that is the probability that an honest  user in the
suspicious set is not removed in the cooperative detection step,  we first
compute the
 probability of false positive before the cooperative detection step at round
 $t$, and denote it by  $P_{fp}^{IS}(t)$. Mathematically,
\begin{equation*}
P_{fp}^{IS}(t)\approx P_{fp}(t-1)\frac{|\CD(t-1)\cap \CD(t)|}{|\CD(t-1)|}.
\end{equation*}
Thus, there are $P_{fp}^{IS}(t)(N-k)$ honest users in the suspicious set if
the detector has $k$ dishonest neighbors. Since $|\CC(t)|$ users are removed
from the suspicious set in the cooperative detection step, we have
\begin{equation*}
P_{CS}(t) = \frac{P_{fp}^{IS}(t)  (N-k) - |\CC(t)| }{P_{fp}^{IS}(t)  (N-k)}.
\end{equation*}

Now probability of false positive after $t$ rounds can be derived as
follows.
\begin{equation}
P_{fp}(t)\approx \frac{P_{fp}(t\!-\!1)\frac{|\CD(t-1)\cap \CD(t)|}{|\CD(t-1)|} (N-k) -|\CC(t)|}
{N-k}.
\label{eq:pfpcooperative}
\end{equation}
If $k \ll N$, then probability of false positive $P_{fp}(t)$ after $t$
rounds can be approximated as
\begin{equation}
P_{fp}(t)\approx \frac{P_{fp}(t-1)\frac{|\CD(t-1)\cap \CD(t)|}{|\CD(t-1)|} N -|\CC(t)|} {N}.
\label{eq:pfpcooperativefinal}
\end{equation}
Note that if $k \ll N$ does not hold,  then  probability of false positive
 in Equation (\ref{eq:pfpcooperativefinal}) is just an
overestimation of Equation (\ref{eq:pfpcooperative}), so it is still
feasible to be used in the termination condition of the complete algorithm.

\section*{Proof of Theorem~\ref{theo:userchurn} in Section~\ref{sec:churn}}

Inspired from the previous analysis, we divide the detection at round $t$
into three steps: (1) the independent detection step, (2) the cooperative
detection step, and (3) the detection step dealing with user churn.
Moreover, probability of false positive after the cooperative detection step
at round $t$ can be derived by Equation (\ref{eq:pfpcooperative}), and we
denote it as $P_{fp}^{CS}(t)$. Since users in $\CNU(t)$ connect with detector $i$ and users in $\CL(t)$
 leave detector $i$ at round $t$, suppose that dishonest users only account for a small
fraction of the population, there are around $N(t-1)-k+|\CNU(t)|-|\CL(t)|$
honest users in the neighboring set after round $t$, where $N(t-1)$ denotes
the number of neighbors of detector $i$ after round $t-1$ and
$N(t)=N(t-1)+|\CNU(t)|-|\CL(t)|$. Moreover, the number of honest users in
the suspicious set after round $t$ is $P_{fp}^{CS}(t) * (N(t-1)-k) +
|\CNU(t)| - |\CL_S(t)|$, so  probability of false positive after $t$ rounds
can be computed via $P_{fp}(t) \approx\frac{P_{fp}^{CS}(t)
* (N(t-1)-k) + |\CNU(t)| - |\CL_S(t)|}{ N(t)-k}$. If $k \ll N(t-1) $ and we substitute
$P_{fp}^{CS}(t)$  with the result in Equation (\ref{eq:pfpcooperative}), we
have
{\small
\begin{equation}
P_{fp}(t)\!\!\approx\!\!\frac{P_{fp}(t\!\!-\!\!1)\frac{|\CD(t\!-\!1)\cap \CD(t)|}{|\CD(t\!-\!
1)|}
 N(t\!-\!1)\!\!-\!\!|\CC(t)| \!\!+\!\! |\CNU(t)|\!\! -\!\! |\CL_S(t)|}
{N(t)}.\nonumber
\end{equation}
}

Again, if $k \ll N(t)$ for all $t$ does not hold, probability of false
positive computed via the above equation  is just overestimated, and it is
still effective to use it to design the termination condition of the
complete algorithm.

\section*{Acknowledgments}

The work of Yongkun Li was supported in part by National Nature Science Foundation of China under Grant No. 61303048,
and the Fundamental Research Funds for the Central Universities under Grant No. WK0110000040.

{\small
\bibliographystyle{abbrv}
\bibliography{pl}
}

\vspace{-0.3in}
\begin{IEEEbiography}[{\includegraphics[width=1in,height=1.25in,clip,keepaspectratio]{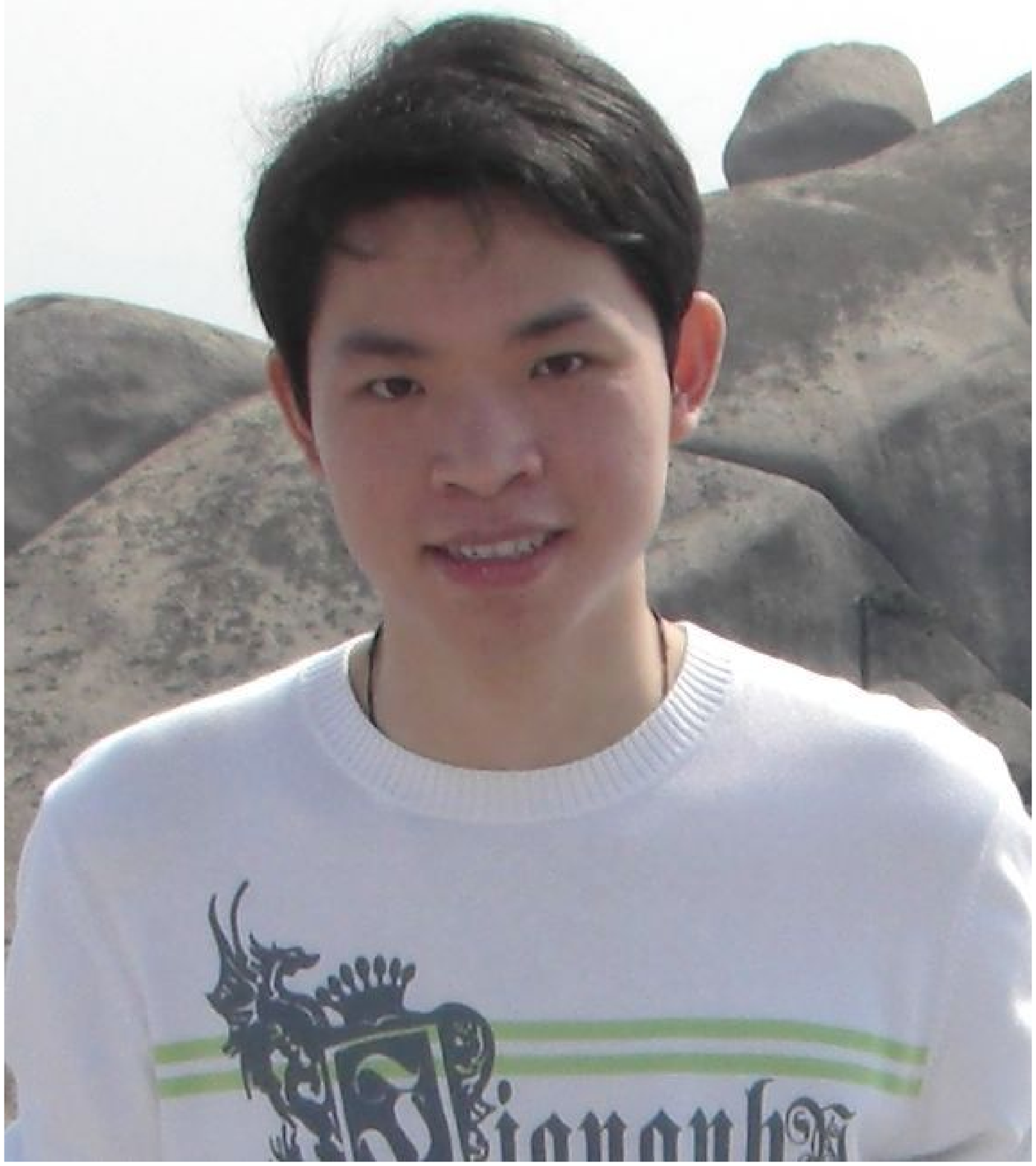}}]{Yongkun Li}
is currently an associate researcher in School of Computer Science and Technology, University of Science and
Technology of China. He
 received the B.Eng. degree in Computer Science from University of
Science and Technology of China in 2008, and the Ph.D. degree in Computer Science and Engineering from The Chinese
University of Hong Kong in 2012. After that, he worked as  a postdoctoral  fellow in Institute of Network Coding at
The Chinese University of Hong Kong. His  research  mainly focuses on performance evaluation of  networking and
storage systems.
\end{IEEEbiography}

\vspace{-0.3in}
\begin{IEEEbiography}[{\includegraphics[width=1in,height=1.25in,clip,keepaspectratio]{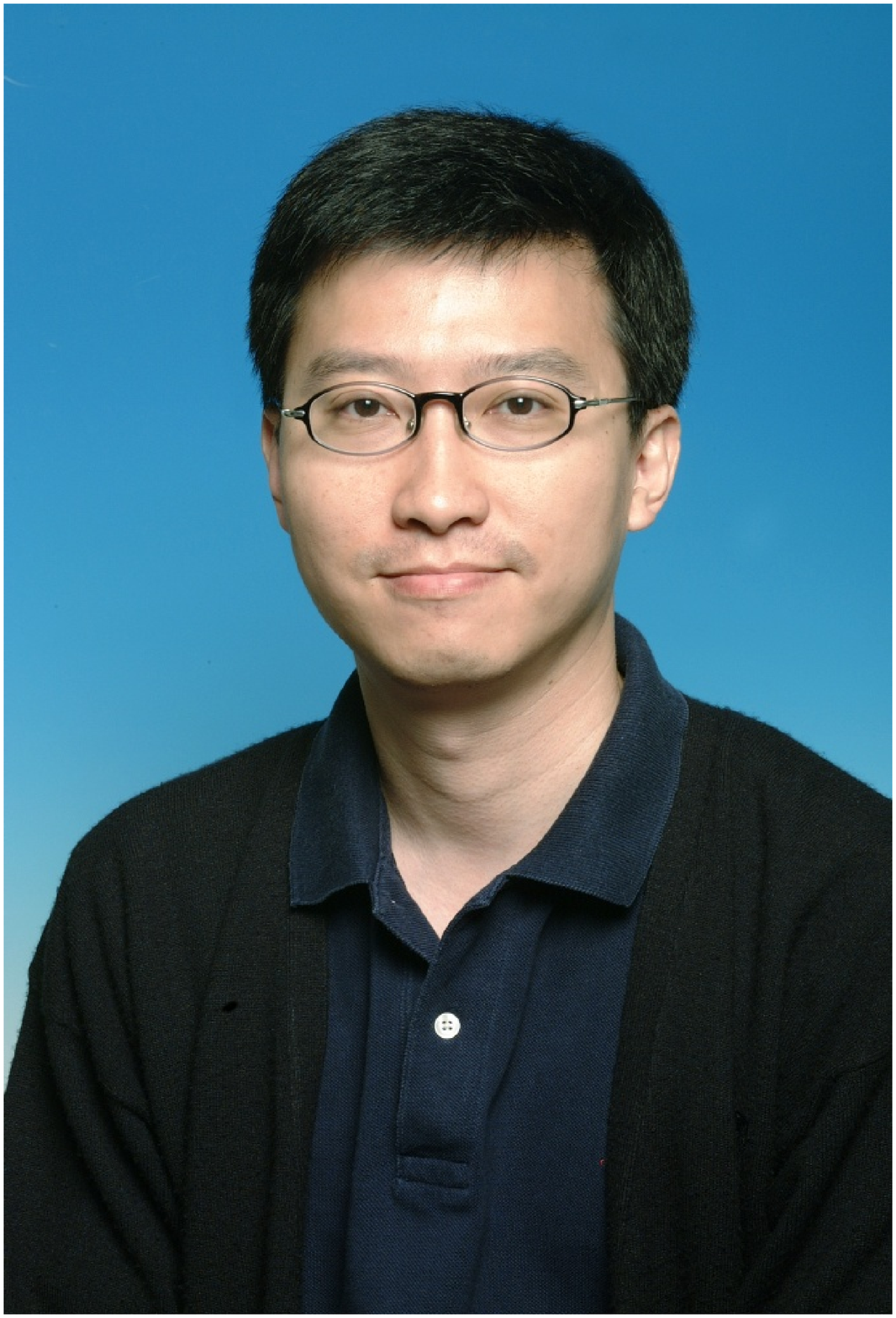}}]{John C. S. Lui } is
currently a professor in the Department of Computer Science \& Engineering at The Chinese University of Hong Kong. He
received his Ph.D. in Computer Science from UCLA. When he was a Ph.D student at UCLA, he worked as a research intern
in the IBM T. J. Watson Research Laboratory. After his graduation, he joined the IBM Almaden Research Laboratory/San
Jose Laboratory and participated in various research and development projects on file systems and parallel I/O
architectures. He later joined the Department of Computer Science and Engineering at The Chinese University of Hong
Kong. John serves as reviewer and panel member for NSF, Canadian Research Council and the National Natural Science
Foundation of China (NSFC). John served as the chairman of the CSE Department from 2005-2011. He serves in the
editorial board of IEEE/ACM Transactions on Networking, IEEE Transactions on Computers, IEEE Transactions on Parallel
and Distributed Systems, Journal of Performance Evaluation and International Journal of Network Security. He received
various departmental teaching awards and the CUHK Vice-Chancellor's Exemplary Teaching Award. He is also a corecipient
of the IFIP WG 7.3 Performance 2005 and IEEE/IFIP NOMS 2006 Best Student Paper Awards. He is an elected member of the
IFIP WG 7.3, Fellow of ACM, Fellow of IEEE and Croucher Senior Research Fellow. His current research interests are in
communication networks, network/system security, network economics, network sciences, cloud computing, large scale
distributed systems and performance evaluation theory.
\end{IEEEbiography}

\end{document}